\documentclass{book}
\usepackage[contrib,lang=british]{ems-book} 
\usepackage{bbm,pgfplots,subcaption,mathtools}
\usepgfplotslibrary{fillbetween}

\newcommand{\real}{\mathbb{R}}

\newcommand{\I}{\mathcal{I}}

\newcommand{\mc}[1]{\mathcal{#1}}
\newcommand{\od}{\text{d}}
\newcommand{\pd}{\partial}

\newcommand{\npderiv}[1]{\frac{\pd}{\pd #1}}
\newcommand{\nsecpderiv}[1]{\frac{\pd^2}{\pd #1 ^2}}

\newcommand{\mean}[1]{\left \langle #1 \right \rangle}

\newcommand{\mb}[1]{\mathbf{#1}}

\begin{document}
\mainmatter

\title{Voter demographics and socio-economic factors in
kinetic models for opinion formation}
\titlemark{Voter demographics and socio-economic factors in
kinetic models for opinion formation}

\emsauthor{1}{Bertram D\"uring}{B.~D\"uring}
\emsauthor{2}{Oliver Wright}{O.~Wright}


\emsaffil{1}{Mathematics Institute, University of Warwick, Zeeman
  Building, Coventry, CV4 7AL, United Kingdom
  \email{bertram.during@warwick.ac.uk}}
\emsaffil{2}{Mathematics Institute, University of Warwick, Zeeman
  Building, Coventry, CV4 7AL, United Kingdom \email{oliver.l.j.wright@warwick.ac.uk}}




\begin{abstract}
Voter demographics and socio-economic factors like age,
sex, ethnicity, education level, income, and other measurable factors
like behaviour in previous elections or referenda are of key
importance in modelling opinion formation dynamics.
Here, we revisit the kinetic opinion formation model from
\textit{D\"uring and Wright} (2022) and compare 
in more detail the influence of different choices of 
characteristic demographic factors and initial conditions. The model
is based on the kinetic opinion formation model by \textit{Toscani} (2006) and the leader-follower model of
\textit{D\"uring et al.} (2009) which leads to a system of
Fokker-Planck-type partial differential equations.
In the run-up to the 2024 general election in the United Kingdom, we consider
in our numerical experiments the situation in the so-called `Red Wall'
in North and Midlands of England and compare our simulation results to other
election forecasts.
\end{abstract}

\makecontribtitle

\section{Introduction}
The importance of democratic elections for societies is undisputed,
and consequently pre-election polling and forecasting election
outcomes based on surveys of political opinion is receiving ever
growing attention. Despite pollsters stressing uncertainties in
their polls and that their surveys represent momentary snapshots of opinions,
weighted averages of polls as functions of time are
presented by third parties in an effort to extract `opinion trends'
and `political momentum'.

Any such efforts are plagued by relatively small sample sizes of  many
polls, e.g. in United Kingdom with its
populations size of around 69 million typical sample sizes in polls
are around 1000-2000. Moreover, there are many methodological issues, e.g.\ the
coverage bias which occurs with the shift from phone to predominantly
online surveys as different groups in society may not engage in online
polls in the same way.

Importantly, even if a pre-election poll
correctly forecasts national level vote shares, winning the
election in plurality (or first-past-the-post) electoral systems
(e.g.\ in UK or US) depends on outcomes at sub-national level. 
It is difficult to forecast outcomes at sub-national level
from classical pre-election polls. This led to repeated (real or perceived) failures of polling. Examples
include the US presidential elections of 2000 and 2016 in which polls correctly
predicting the winner of the popular vote, but not of the
Electoral College, and the 2015 general elections in the United Kingdom where almost
every poll predicted a hung parliament, i.e. a parliament with no
clear majority, but in the elections the Conservative party won with a
strong majority.

In recent years, so-called multilevel regression and
post-stratification (MRP) methods \cite{park2004bayesian} have emerged as a promising tool to
improve outcome forecasts at sub-national level. They rely on larger
sample sizes and on recording demographic and socio-economic
characteristics like age, sex, ethnicity, education level, income and
information on previous votes in elections or referenda, in addition to
the participants' voting intentions.
A regression is performed to determine how likely it is that an individual with a certain
combination of demographic and socio-economic characteristics votes for a particular party.
In a post-stratification step, census and other publicly available data are used to determine the
composition of the voter population at sub-national level, allowing to
more accurately forecast the sub-national and with it also the
national election outcome.

Motivated by the success of MRP methods, we have proposed and analysed a kinetic model for opinion formation which takes
into account demographic and socio-economic factors like age, sex, ethnicity, education level,
income and other measurable factors in \cite{bib:During:polling}.
Kinetic models for opinion formation have been considered by many authors,
see e.g.\
\cite{bib:Toscani:kineticmodel,bib:During:strongleaders,bertotti2008discrete,motsch2014heterophilious,zbMATH06350891,bib:During:inhomogeneous,albi2016opinion,bib:ToscaniPareschi:ims,toscani2018opinion,albi2015optimal,albi2016recent,zbMATH06665924,zbMATH06553232,zbMATH06684855,zbMATH06684856,zbMATH06712347,zbMATH06735322,zbMATH06875734,zbMATH06892489,zbMATH06977672,zbMATH07205672,zbMATH06944946,albi2019boltzmann,zbMATH07146312,burger2021network,during2024breaking,zanella2023kinetic,albi2024data}.
Building on mathematical approaches from kinetic theory they consider
the passage from interacting many-agent systems to Boltzmann-type
equations. As reminiscent of binary collisions in gas theory, one considers the
process of compromise between two individuals. Similar models are used
in the modelling of wealth and income distributions which show Pareto
tails, cf.\ \cite{bib:During:economykineticmodel} and the references
therein.

In this chapter we revisit the model from \cite{bib:During:polling} and compare
in more detail the influence of different choices of 
characteristic demographic factors and initial conditions.
With an interest in the upcoming 2024 general election in the United Kingdom, we consider
in our numerical experiments the situation in the so-called `Red Wall'
in North and Midlands of England and compare our simulation results to other
election forecasts.

The remainder of this chapter is organised as follows. In the next
section, we recall the kinetic
opinion formation model from \cite{bib:During:polling} which takes into account demographic and socio-economic factors in
the compromise process of the kinetic opinion formation model. In the
quasi-invariant opinion limit this leads to a system of
Fokker–Planck-type equations. In Section~\ref{Pfunction}, we discuss the data-driven
compromise functions which are used to introduce a bias into the
consensus formation dynamics, depending on voter intention data. We
briefly discuss the numerical discretisation of the Fokker-Planck
system in Section~\ref{Numerics}, before we turn to an application of the
model in the upcoming 2024 UK general elections in Section~\ref{GE2024}. We show simulation
results, discuss the choice of initial conditions and demographic
factors, and compare our results to other predicted results for the
upcoming election. Since the 2024 UK general election has been called
for 4 July 2024 and this work was completed before this date we do not
compare our results with the recorded election results.

\section{Kinetic opinion formation model of D\"uring and Wright (2022)}
We recall our kinetic opinion formation model from
\cite{bib:During:polling} in which we propose a data-driven
modification leader-follower model which takes into account voter
demographics, and socio-economic and other factors.
It is based on the leader-follower model of {\em D\"uring et al.}
\cite{bib:During:strongleaders}. Opinion is a continuous independent
variable, $w$, in an interval $\mathcal{I}=[-1,1]$, the other
independent variable is time $t\ge0$. The population is
split into $N$ follower species and a leader species, denoted by
$f_i(w, t)$, $i=1,2,...,N$, and $f_L(w,t)$, respectively. Each
follower species is associated with one combination of $n$
``characteristics", $c_{i1}, c_{i2}, ... , c_{in}$, each of which is
chosen from a potential set of mutually exclusive groups.
These characteristics include demographic and socio-economic factors
like age, sex, ethnicity, education level, income and other measurable
factors like behaviour in previous elections or referenda.  The leader
species is formed of  strong opinion leaders, as in
\cite{bib:During:strongleaders}, that is assertive individuals able to
influence followers but withstand being influenced themselves.

The dynamics of the microscopic many-agent system are characterised by binary
 interactions, similar as in
 \cite{bib:Toscani:kineticmodel,bib:During:strongleaders}. In the present
 situation there are three types of possible interactions with $w_L^*,
 v_L^*$ and $w_L, v_L$ as the post- and pre-interaction leader
 opinions, respectively, and $w^*, v^*$ and $w, v$ as the post- and pre-interaction opinions of the follower species, respectively:
    \begin{enumerate}[(a)]
        \item $L$ and $L$ interaction:
            \begin{equation} \label{int:LL}   
                \begin{split}
                    w_L^* &= w_L+ \gamma_L P_{LL}(w_L,v_L) (v_L-w_L) + D_L(w_L) \eta_L, \\ 
                    v_L^* &= v_L+ \gamma_L P_{LL}(v_L,w) (w_L-v_L) + D_L(v_L) \eta_L;
                \end{split}
            \end{equation}
        \item $i$ and $L$ interaction, $i = 1,2,...,N$:
            \begin{equation} \label{int:cL}   
                \begin{split}
                    w^* &= w + \gamma_i P_{iL}(w, v_L) (v_L-w) + D(w) \eta_i, \\ 
                    v_L^* &= v_L;
                \end{split}
            \end{equation}
        \item $i$ and $j$ interaction, $i,j = 1,2,...,N$:
            \begin{equation} \label{int:cc} 
                \begin{split}
                w^* &= w + \gamma_i P_{ij}(w,v) (v-w) + D(w) \eta_i, \\ 
                v^* &= v + \gamma_j P_{ji}(v,w) (w-v) + D(v) \eta_j.
                \end{split}
            \end{equation}
    \end{enumerate}
    
    In the interactions above, $\gamma_\nu  \in (0,{1}/{2})$, for $\nu = 1, 2,... N, L$, denotes compromise parameters which govern the ``speed'' of attraction of two different opinions, the functions $P_{ij}(\cdot, \cdot)$ and $D(\cdot)$ model the local relevance of compromise and self-thinking for a given opinion, respectively, and $\eta_\nu$ are the ``self-thinking" random variables, which model how the opinion of an individual changes independently from other individuals.

  To ensure that $w^*$ and $v^*$ remain in $\I$, the functions
  $P_{ij}$ and $D, D_L$ functions have to be chosen accordingly, see
  \cite{bib:During:polling}.
  The important novelty of this model from \cite{bib:During:polling} is the
  introduction of data-driven functions $P_{ij}$ which are based on
  polling data, explained in more detail in Section~\ref{Pfunction}
  below. A usual choice for $D$ which we adopt also here is $D(x) = (1-x^2)^\alpha$, for some $\alpha>0$. We choose $D_L$ similarly and for both $\alpha = 2$, unless otherwise stated.

    Finally, we now discuss the random variables $\eta_\nu$. Let us define a set of probability measures in the following way, for some $\delta>0$: 
    \begin{align*}
    \mc{M}_{2+\delta} = \bigg\{ \mathbb{P} : \mathbb{P}& \text{ is a probability measure and, } \\
    &\mean{|\eta|^\alpha} =  \int_\I |\eta|^\alpha \ \od \mathbb{P}(\eta)<\infty, \forall \ \alpha \le 2+\delta  \bigg\}.
    \end{align*}
    We choose a probability density $\Theta \in \mc{M}_{2+\delta}$ and we assume that this density is obtained from a random variable $Y$ with the properties,  $\mean{Y} = \int_\I Y \ \od \Theta (Y) = 0$, $\mean{Y^2} = \int_\I Y^2 \ \od \Theta (Y) = 1$, then we choose $\eta_\nu$ with the property that $\mean{|\eta_\nu|^p} = \mean{|Y\sigma_\nu|^p} $ for all $ \nu = 1, 2,..., N,L$ and $p \in [0,2+\delta]$, with $0<\sigma_\nu \ll  1$. 

    In the quasi-invariant limit \cite{bib:Toscani:kineticmodel,
        bib:During:strongleaders} $\gamma_i,
    \sigma_i \rightarrow 0$ with ${\sigma_i^2}/{\gamma_i} =
    \lambda_i$ fixed, we obtain the following Fokker-Planck-type
    system \cite{bib:During:polling},
    \begin{equation}\label{Fokker:Planck:chap2}
        \begin{split}
            \npderiv{t}  (f_i(w,t)) =&  \npderiv{w}\Bigg[ \bigg(\frac{1}{2\tau_{iL}} \mc{K}_{iL}[f_L](w,t) + \sum_{j=1}^N \frac{1}{2\tau_{ij}} \mc{K}_{ij}[f_j](w, t)\bigg)  f_i(w,t) \Bigg] \\
            & +\sum_{j=1}^N \frac{\alpha_{ij}}{2\tau_{ij}} \npderiv{w}\bigg( \mc{K}_{ij}[f_i](w,t) f_j(w,t)  \bigg) \\
            & + \left( \frac{\lambda_i M[f_L](t)}{4\tau_{iL}} + \sum_{j=1}^N \frac{\lambda_i M[f_j](t)}{4\tau_{ij}} \right) \nsecpderiv{w}\bigg( D^2(w) f_i(w,t) \bigg) \\
            & + \sum_{j=1}^N \frac{\beta_{ij} \lambda_i M[f_i](t)}{4\tau_{ij}} \nsecpderiv{w} \bigg( D^2(w)  f_j(w,t) \bigg) \text{, for } i = 1,2,...,N, \\
            \npderiv{s}  ( f_L(w,t)) =&  \frac{\alpha_{iL}}{\tau_{L}} \npderiv{w}\Bigg(  \mc{K}_{LL}[f_{L}](w,t)   f_L(w,t) \Bigg)  \\
            & +\frac{\beta_{iL} \lambda_L M[f_L](t)}{2\tau_{L}}  \nsecpderiv{w}\bigg( D_L^2(w) f_L(w,t) \bigg),
        \end{split} 
    \end{equation}
    where, with $\xi, \mu, \nu = 1,2,...,N,L$,
    \begin{equation*}
        \mc{K}_{\mu \nu}[f_\xi](w,t) := \int_\I P_{\mu \nu}(w,v)(v-w) f_\xi(v,t) \ \od v, \ M[f_{\xi}](t) := \int_\I f_\xi(w,t) \ \od w.
      \end{equation*}

 \section{Compromise function from demographic voter
   intention data}\label{Pfunction}

In \cite{bib:During:polling} a data-driven approach to include
demographic and other factors into the compromise function has been introduced whereas previously, the compromise function $P$ has been chosen
     as a localisation function, see e.g.\ \cite{bib:During:inhomogeneous,
       bib:During:strongleaders, bib:Toscani:kineticmodel}.
To this end, data from voter intention polls made before the election
are employed.

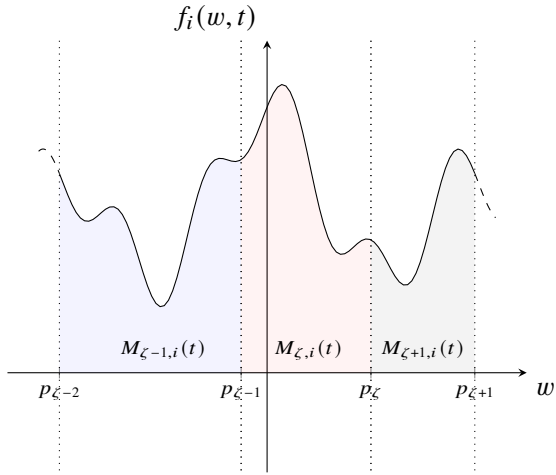
\begin{figure}[h]
    \begin{center}
        \begin{tikzpicture}
    	    \begin{axis}[
        		axis x line=center,
    	    	axis y line=center,
    		    xlabel={$w$},
    			ylabel={$f_i(w,t)$},
    			xlabel style={below right},
    			ylabel style={above left},
    			xticklabels ={,,},
    			yticklabels = {,,},
    			xtick={-1, -0.125, 0.5, 1},ytick={\empty},
    			xmin=-1.25,
    			xmax=1.25,
    			ymin=-0.3,
    			ymax=1.0]
    
    			\addplot [name path=f,mark=none,domain=-1:1,samples=100,yticklabels=\empty] {0.1*cos(deg(x*3.141593))+ 0.2*cos(deg(2*x*3.141593)) + 0.1*sin(deg(5*x*3.141593))+0.5};
                \addplot [dashed, mark=none,domain=1:1.1,samples=100,yticklabels=\empty] {0.1*cos(deg(x*3.141593))+ 0.2*cos(deg(2*x*3.141593)) + 0.1*sin(deg(5*x*3.141593))+0.5};
                \addplot [dashed, mark=none,domain=-1.1:-1,samples=100,yticklabels=\empty] {0.1*cos(deg(x*3.141593))+ 0.2*cos(deg(2*x*3.141593)) + 0.1*sin(deg(5*x*3.141593))+0.5};
    
            \path[name path=axis1] (axis cs:-1,0) -- (axis cs:-0.125,0);
   
    			\addplot[mark='',only marks] coordinates{(0.-1,-0.01)}node[anchor=north]{\scriptsize{$p_{\zeta-2}$}};
    			\addplot[mark='',only marks] coordinates{(0.-0.125,-0.01)}node[anchor=north]{\scriptsize{$p_{\zeta-1}$}};				
                \addplot [fill=blue, fill opacity=0.05] fill between[of=f and axis1, soft clip={domain=-1:-0.125},];
                \addplot[mark='',only marks] coordinates{(-0.5,0.01)}node[anchor=south]{\scriptsize{$M_{\zeta-1,i}(t)$}};
        		\draw[dotted] (axis cs:-1,1.25) -- (axis cs:-1,-1.25);
    			\draw[dotted] (axis cs:-0.125,1.25) -- (axis cs:-0.125,-1.25);
    
            \path[name path=axis2] (axis cs:-0.125,0) -- (axis cs:0.5,0);
    
    		\addplot[mark='',only marks] coordinates{(0.5,-0.01)}node[anchor=north]{\scriptsize{$p_\zeta$}};
    			\addplot [fill=red, fill opacity=0.05] fill between[of=f and axis2, soft clip={domain=-0.125:0.5},];
                \addplot[mark='',only marks] coordinates{(0.2,0.01)}node[anchor=south]{\scriptsize{$M_{\zeta, i}(t)$}};
\draw[dotted] (axis cs:-0.125,1.25) -- (axis cs:-0.125,-1.25);
    			\draw[dotted] (axis cs:0.5,1.25) -- (axis cs:0.5,-1.25);
    
            \path[name path=axis3] (axis cs:0.5,0) -- (axis cs:1,0);
    			\addplot[mark='',only marks] coordinates{(1,-0.01)}node[anchor=north]{\scriptsize{$p_{\zeta+1}$}};
    			\addplot [fill=black, fill opacity=0.05] fill between[of=f and axis3, soft clip={domain=0.5:1},];
                \addplot[mark='',only marks] coordinates{(0.75,0.01)}node[anchor=south]{\scriptsize{$M_{\zeta +1, i}(t)$}};
    			\draw[dotted] (axis cs:0.5,1.25) -- (axis cs:0.5,-1.25);
    			\draw[dotted] (axis cs:1,1.25) -- (axis cs:1,-1.25);
		    \end{axis}
        \end{tikzpicture}
    \end{center}
    \caption{Schematic representation of influence masses for a given density function, $f_i(w,t)$, at some time $t$.} \label{fig:inflmass}
    \end{figure}

Let there be political parties, $\Pi_1, ...,
  \Pi_\mc{P}$, and $p_0 <
  p_1<...<p_{\mc{P}-1}< p_{\mc{P}} $ be a partition of $\I$, then we
  associate the mass of the leader species in an influence interval $\I_\zeta =
  [p_{\zeta-1}, p_\zeta]$ with a political party standing in the
  election.  The fractional masses of follower species contained within
  an influence interval are assumed to vote for the controlling party,
  and give rise to influence masses,
    $$M_{\zeta, i}(t) = \int_{\I_\zeta} f_i(w,t) \ \od w$$
    for every $\zeta = 1,...,\mc{P}$. Figure~\ref{fig:inflmass} shows
    a schematic representation.

    Typical voter intention data is organised in tables by
    characteristics, with parties on the vertical axis and the groups
    (e.g. leave and remain for 2016 Brexit Vote, 18-24 year-old
    etc. for age) on the horizontal axis. A table for an example
    characteristic $k$ is displayed in
    Table~\ref{tab:PFunction:examplechar} and the actual table from
    \cite{bib:YouGov:voterintention2024} for 2024 voter intention data
    relating to their 2016 Brexit referendum vote is displayed in Table~\ref{tab:PFunction:brexitchar}.

    \begin{table}[h] 
    \centering
    \caption{Example of how the characteristic data is presented.}
    \renewcommand{\arraystretch}{1.2}
        \begin{tabular}{c|cccc}
            Characteristic $k$ & Group 1 & Group 2 & $\cdots$ & Group $m$\\
            \hline
            $\Pi_1$ & $x_{1,1,k}$ & $x_{1,2,k}$ & $\cdots$ & $x_{1,m,k}$ \\
            $\Pi_2$  & $x_{2,1,k}$ & $x_{2,2,k}$ & $\cdots$ & $x_{2,m,k}$ \\
            $\vdots$ &  $\vdots$ & $\vdots$  & $\ddots$ & $\vdots$  \\
            $\Pi_\mc{P}$  & $x_{\mc{P},1,k}$ & $x_{\mc{P},2,k}$ & $\cdots$ & $x_{\mc{P},m,k}$ 
        \end{tabular}

        \label{tab:PFunction:examplechar}
    \end{table}

    \begin{table}[h]
    \centering
    \caption{The 2024 voter intention data relating to their 2016
      Brexit referendum vote from \cite{bib:YouGov:voterintention2024}.}
    \renewcommand{\arraystretch}{1.2}
        \begin{tabular}{l|cc}
            Brexit Vote & Leave & Remain \\
            \hline
            Labour & 0.17 & 0.51 \\ 
            Green & 0.03 & 0.08 \\
            LibDems & 0.07 & 0.19 \\
            Other & 0.05 & 0.08 \\
            Conservative &0.35 & 0.11 \\
            Brexit & 0.33 & 0.03 
        \end{tabular}
        \label{tab:PFunction:brexitchar}
    \end{table}

    For each binary interaction \eqref{int:LL}--\eqref{int:cc} we can
    find the influence intervals $\I_\zeta$ to which $w,v$ belong to,
    assume that $w\in \I_\mu$ and $v\in \I_\nu$, for some $\mu, \nu =
    1,...,\mc{P}$. We look up the relevant values in the tables to get
    the values of $x_{\mu,i,k}$ and $x_{\nu,j,k}$.
    
    Let $w,v$ be the opinions of two individuals in follower species $i$ and $j$, respectively, with associated characteristic groups $c_{ik}, c_{jk}$ for all $k=1,...,n$, we use the following compromise function decomposition:
        \begin{equation}
            P_{ij}(w,v) = P_{loc}(w, v) \prod_{k=1}^n P_{char}(w,v,c_{i,k}, c_{j,k}).
        \end{equation}
    We consider the localisation function,
    \begin{equation*}
        P_{loc}(w,v) = \mb{1}_{[-r,r]}(w-v),
    \end{equation*}
    for some $r>0$ where $\mb{1}_{A}(x)$ denotes the characteristic
    function for the set $A$.

    For the leader species interactions $P_{iL} (w,v) = P_{LL}(w,v) = P_{loc}(w,v)
    $. For follower species interactions, we use the characteristic
    comparison function
\begin{equation}
        P_{char}(w,v,c_{i,k}, c_{j,k}) = \sum^\mc{P}_{\mu, \nu = 1} \mb{1}_{[p_{\mu-1}, p_\mu]}(w) \mb{1}_{[p_{\nu-1}, p_\nu]}(v)  F(x_{\mu,i,k}-x_{\nu,j,k}),\label{eq:ourP}
      \end{equation}
 with $$F(x)=\bigg(1-\left(\frac{1}{4}(2+x)\right)^a\bigg)^b,$$ and $a,b > 0$
    as in \cite{bib:During:polling}. 
     The characteristic comparison function $P_{char}$ models the idea
     of net movement towards more popular opinions, as indicated by
     voter intention data from pre-election polls. Individuals with
     current popular opinions move slower towards unpopular opinions
     than individuals with current unpopular opinions move towards
     more popular opinions. The argument of the function,
     $x_{\mu,i,k} - x_{\nu,j,k}$, gives a large, positive value when
     the party, in whose influence the individual resides, is more
     popular (i.e $x_{\mu,i,k}$ dominates) and a large negative value
     when the party, in whose influence the other individual resides,
     is more popular (i.e $x_{\nu,j,k}v$ dominates). When both parties
     are equally popular, this value will be zero and therefore will
     not have a bias towards one party or the other. Therefore, the difference $x_{\mu,i,k} - x_{\nu,j,k}$ is representative of the popularity of individuals' opinion.

      \section{Numerical experiments} \label{Numerics}
      For purposes of the discretisation of the Fokker-Plank system
    \eqref{Fokker:Planck:chap2}  we employ a finite element method. 
Let $(f,g)$ denote the $L^2$ inner product of the function $f$ and $g$, $(f,g)=\int_\I f(w)\cdot g(w) \ \od w.$
We have chosen the values of $a=2$ and $b=8$ in the $P_{char}$
function since these have offered the best results, $\alpha_{i} = \beta_{i} = 1$ for all $i = 1, 2, ..., N, L$. We construct our initial densities from available polling data and from population values for the specific region we wish to run simulations in. We do this by first finding the influence mass of the Party $\Pi_\zeta$ for a given species $i$, multiplying the relevant columns of our data tables, which, for a species whose members have characteristic $k$ and are in Group $\xi$, is the $\xi$-th column from Table~\ref{tab:PFunction:examplechar} and the relevant column from every characteristics table, element-wise. We will call the elements of the resulting column $X_{\zeta,i}$. Recall that by definition 
    $$M_i = \sum_{\zeta}M_{\zeta,i},$$
    where we recall that $M_{\zeta,i}$ are the influence masses from Section~\ref{Pfunction}. Given some distribution of mass between the follower species $M_i$, we can compute the initial influence masses for the parties in  
    $$M_{\zeta, i} = \frac{v_\zeta X_{\zeta,i}}{\sum_{\zeta} v_\zeta X_{\zeta, i}}  M_i,$$
    with $v_\zeta$ being the percentage of votes allocated to $\Pi_\zeta$ in some test election, either through the recorded results of some previous election or some pre-election polling data. We define the initial densities of the species as a weighted sum of characteristic functions in the following way
    $$f_i(w,0) = \sum_\zeta M_{\zeta,i} \mathbbm{1}_{w\in \I_\zeta},$$
    for all $\zeta = 1,2,...,N$.
    
    The initial density of the leader species is chosen as a weighted sum of delta distributions,
    \begin{equation} \label{eq:initiallead}
        f_L(w,0) = \sum_{k=1}^6 p_k\delta(w-L_k),
    \end{equation}
    with $p_k$ representing the percentage of the vote that each party
    attained in this, or another, test election, and $L_k$
    representing the party locations. In all the simulations presented
    here, we set the mass of the leader species to have 5\% the total
    mass of the follower species, in a similar way to \textit{D\"uring
      et al} \cite{bib:During:strongleaders}.

    We approximated the delta functions in $f_L(w,0)$ with the scaled mollifier functions
    \begin{equation*}
        \phi(x) = \begin{dcases}
        \exp\left(-\frac{({x}/{\varepsilon})^2}{1-({x}/{\varepsilon})^2}\right), & \text{for } x \in [-1,1], \\
        0, & \text{otherwise},
                   \end{dcases}
    \end{equation*}
    with scale factor $\varepsilon \ll 1$.

We discretise our time axis with $A=300$ nodal points, and space axis with $B=100$ nodal points. We write the nodal values of our functions $f_i(w_k,t^n) = f_{i,k}^n$ and the vector of every spatial nodal value, at time $t^n>0$, as $\mb{f}_i^n$. We then use a semi-implicit Euler method to discretise our coupled system of non-linear, non-local PDEs in time and an FEM for discretisation in space, with quadratic, Lagrangian basis functions $b_k \in C_c(\I;\real)$ for $k = 1,2,...,B $, which leads to the following linear system:
    \begin{equation} \label{FPM:discrete}
            \sum_{i=1}^N \bigg((\mb{M} + \mb{U}_{i} - \mb{C}^n_{Li}) \mb{f}^{n+1}_i + \sum_{j=1}^N (\mb{V}_{ij} - \mb{T}_{ij}^n) \mb{f}_j^{n+1}\bigg) = \sum_{i=1}^N \mb{M} \mb{f}^n_i,
    \end{equation}
     where $\mb{M} = (b_k,b_l)_{(k,l)}$ is the mass matrix, $\mb{C}^n_{Li}$ and $\mb{T}_{ij}^n$ are the discretised transport matrices,
given by
    \begin{align*}
    \mb{C}^n_{Li} &= \Delta t  \left(\left[\frac{1}{2\tau_{iL}}K_{L,iL}^n + \sum_{j=1}^N \frac{1}{2\tau_{ij} }K_{j,ij}^n\right]b_k, b_l'\right)_{(k,l)}, \\ 
    \mb{T}_{ij}^n &= \Delta t \left(\frac{\alpha_{ij}}{2\tau_{ij}} K_{i,ij}^n b_k, b_l'\right)_{(k,l)},    
    \end{align*} 
    with $K_{L,iL}^n(w)$ and $K_{m,ij}^n(w)$ as the midpoint rule approximation of the convolution operators, $\mc{K}_{ij}[f_m] (w, t^n)$ evaluated at the previous time step, in the transport terms of the Fokker-Planck equations and $\mb{U}_i$ and $\mb{V}_{ij}$ are the discretised Laplacian term matrices,
    \begin{align*}
        \mb{U}_i &= \Delta t \left(\left[\frac{\lambda_i M_L}{4\tau_{iL}} + \sum_{j=1}^N \frac{\lambda_i M_j}{4\tau_{ij}}\right]D^2 b_k',b_l'\right)_{(k,l)},\\
        \mb{V}_{ij} &= \Delta t \left(\frac{\lambda_i M_j \beta^2_{ij}}{4\tau_{ij}\alpha_{ij}}D^2 b_k', b_l'\right)_{(k,l)}.
    \end{align*}
\section{Application to the 2024 General Election in the United Kingdom}\label{GE2024}

In this section, we consider the upcoming 2024 general election in the
United Kingdom, focusing on Nottinghamshire as part of the so-called
Red Wall (a group
of constituencies in the North and Midlands of the UK traditionally
dominated by Labour), as in
\cite{bib:During:polling}. When comparing with the earlier results, note that in the time between this work and
the one presented in \cite{bib:During:polling}, the Brexit party has
been renamed to Reform UK and we will use this name throughout this
work. Also the constituency boundaries have been modified and
subsequently the constituencies of Nottingham North and Sherwood have
been renamed to Nottingham North and Kimberley and Sherwood Forest,
respectively.

We present here a blanket of possible results that could arise from these models. Namely, we will consider the use of different demographics when defining the follower species, the use of universal initial conditions and investigate the role of the maximum time used in the model. Finally, we compare the results from these investigations with other predictions of the election.

Unless otherwise stated, we again use the parameters $\tau_{iL} =
0.05$, $\tau_{ij} = 2.5$, $\lambda = 0.033$, $\Delta t = 0.1$ and $r =
0.85$. Initial masses and regionalised data are produced similarly to
\cite{bib:During:polling} with the exception that we use a more recent
poll from YouGov as our voting intention data
\cite{bib:YouGov:voterintention2024}. Other data used are
2011 Census data \cite{bib:ONS:2011census},  results of
the 2017, 2019 UK general elections 
\cite{bib:HOC:2017GE,bib:HOC:2019GE}, and results of the 2016 EU referendum \cite{bib:ElectoralComission:2016Brexit}.

\subsection{Investigating the number of time steps}

In this experiment, we explore the effect of changing the final time of the model. In previous examples, we have left the model to run for very large times, in order to reach the equilibrium state. However this may not be a particularly realistic method, and may lead to over-fitting the model to the original data. Therefore in this experiment, we run the model for both $A = 10$ and $A=100$ to investigate this difference. Note that although the time steps are different in each model, the change in time between time steps $\Delta t$ remains at $0.1$ -- that is the final time will be $1$ in the first model and $10$ in the second.

        \begin{figure}[h]
    	    \centering
    	    \begin{subfigure}[b]{0.45\linewidth} 
    	        \centering
    		    \includegraphics[scale=0.35]{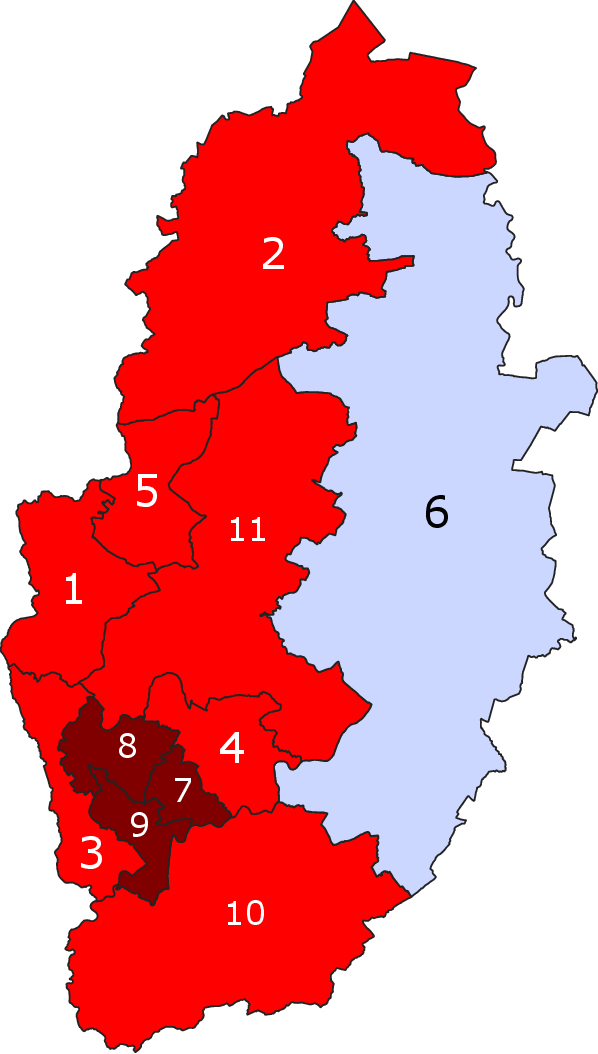}
        		\caption{Predicted Result from the kinetic model with $A=10$.}
        		\label{fig:4.3analysis:Not:2024T10}
    	    \end{subfigure}
        \hspace{0.04cm}
        	\begin{subfigure}[b]{0.45\linewidth} 
        	\centering
    	    	\includegraphics[scale=0.35]{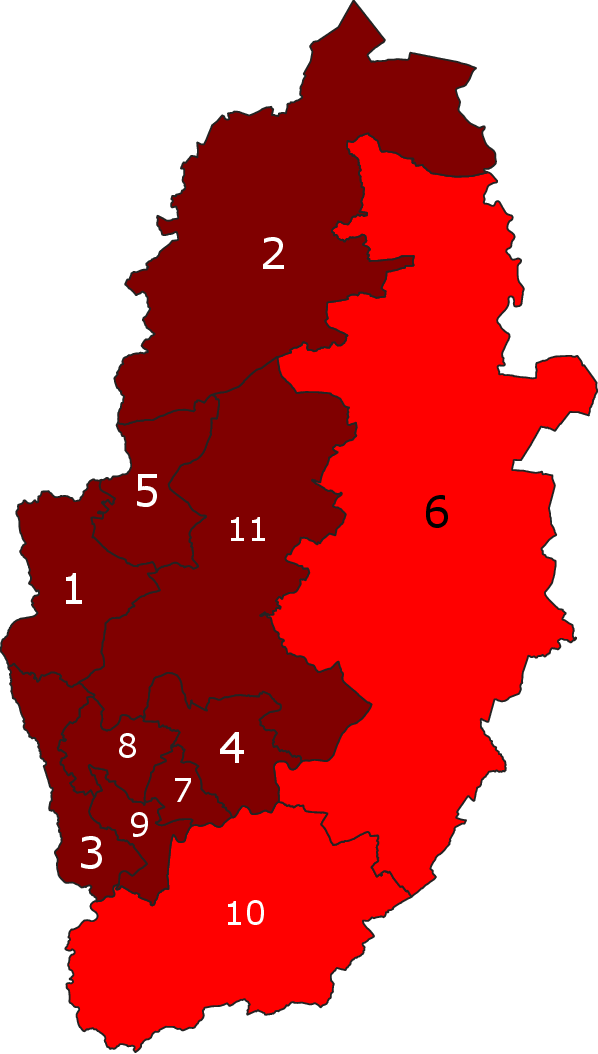}
    		    \caption{Predicted Result from the kinetic model with $A=100$.}
    		    \label{fig:4.3analysis:Not:2024T100}
    	    \end{subfigure}
        	\caption{Predicted Results of the General Elections in
                  2024 in Nottinghamshire with different total
                  timesteps. Light colours (Labour in red,
                  Conservative in blue) are where the total popular
                  vote for the elected MP was below 50\% and the dark colours for greater than 50\%.}
        	\label{fig:4.3analysis:timeNot}
        \end{figure}
The results are presented in Figure~\ref{fig:4.3analysis:timeNot}, with
colour-coding of each constituency, red if a Labour MP was elected,
blue if a Conservative MP was elected. The light colours are where the
total popular vote for the elected MP was below 50\% and the dark
colours for greater than 50\%. The results show a stark difference in final winning percentage, but not the winning party in all constituencies, except (6) Newark, however the final results in the model  Figure~\ref{fig:4.3analysis:Not:2024T10}, the difference between Labour and Conservative are very small compared to the other constituencies. It would appear that using a much higher final time for the model does lead to some over-fitting but the strength of this effect and the optimal stopping time are left as areas of further study. In the following models, we take the number of temporal nodal points to be $A=10$.

\subsection{Investigating the choice of initial condition}

\begin{figure}
	    \begin{subfigure}[b]{0.45\linewidth}
		    \includegraphics[width=\textwidth]{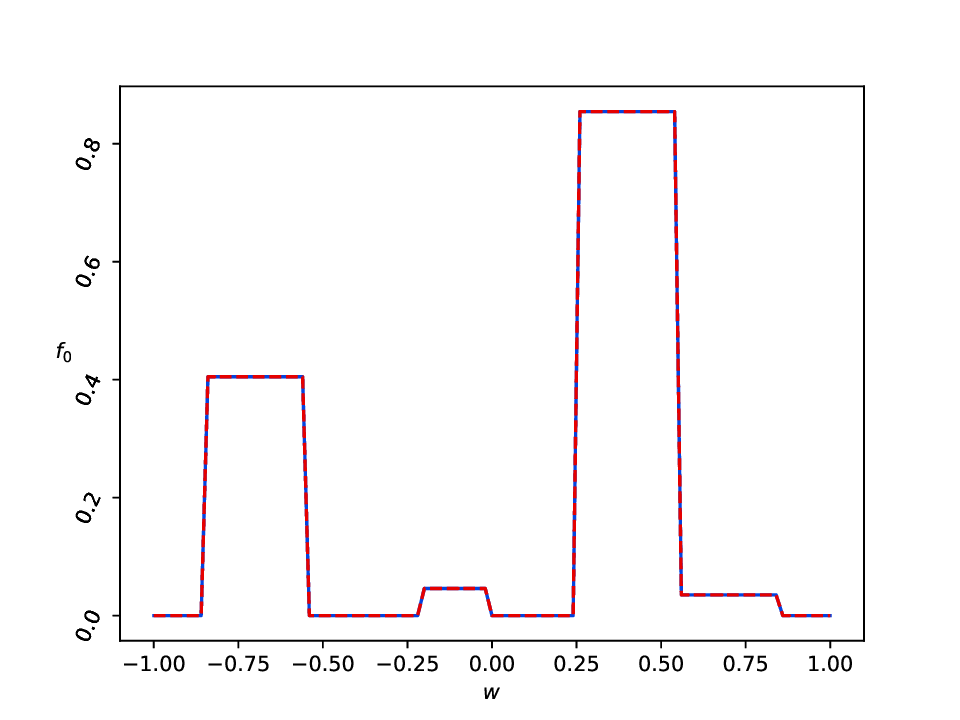}
		    \caption{18--24 year-old}
                  \end{subfigure}%
                       \hfill
                  \begin{subfigure}[b]{0.45\linewidth}
		    \includegraphics[width=\textwidth]{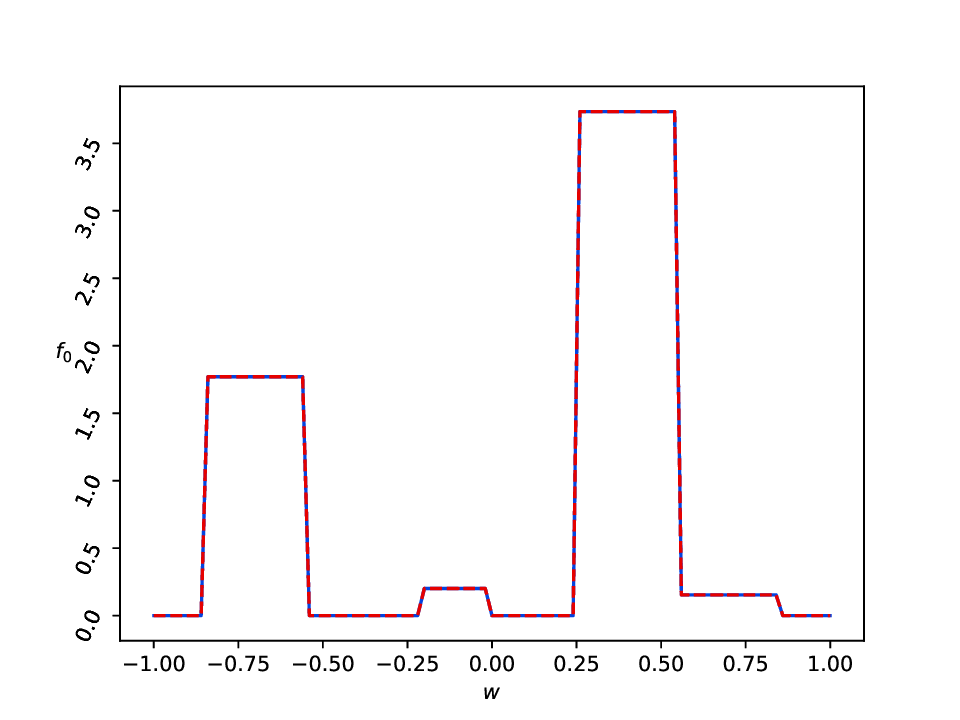}
		    \caption{25--49 year-old}
	    \end{subfigure} 
    	\begin{subfigure}[b]{0.45\linewidth}
	    	\includegraphics[width=\textwidth]{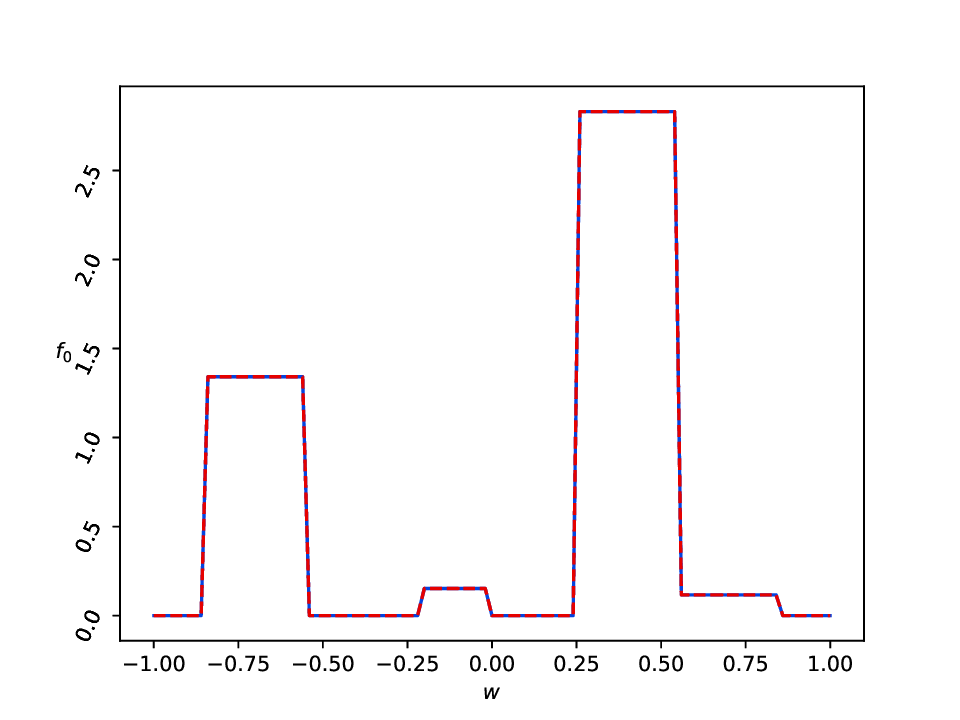}
		    \caption{50--64 year-old}
                  \end{subfigure} %
                       \hfill
	    \begin{subfigure}[b]{0.45\linewidth}
		    \includegraphics[width=\textwidth]{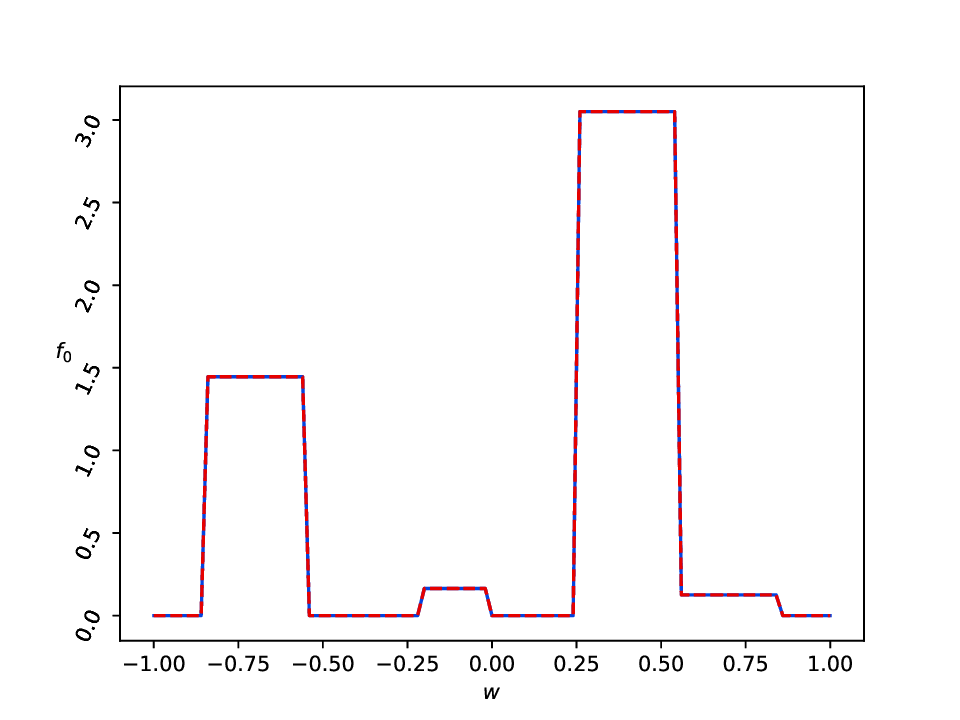}
            \caption{65+ year-old}
	    \end{subfigure}
	    \caption{Universal initial densities for the eight follower species in the constituency of Newark from the 2017 General Election Results. Each graph shows the initial density for the leave (dashed red) and the remain (solid blue) characteristic of an age group.}
	    \label{fig:4.3analysis:newunivICs}
    \end{figure}

\begin{figure}
	    \begin{subfigure}[b]{0.45\linewidth}
		    \includegraphics[width=\textwidth]{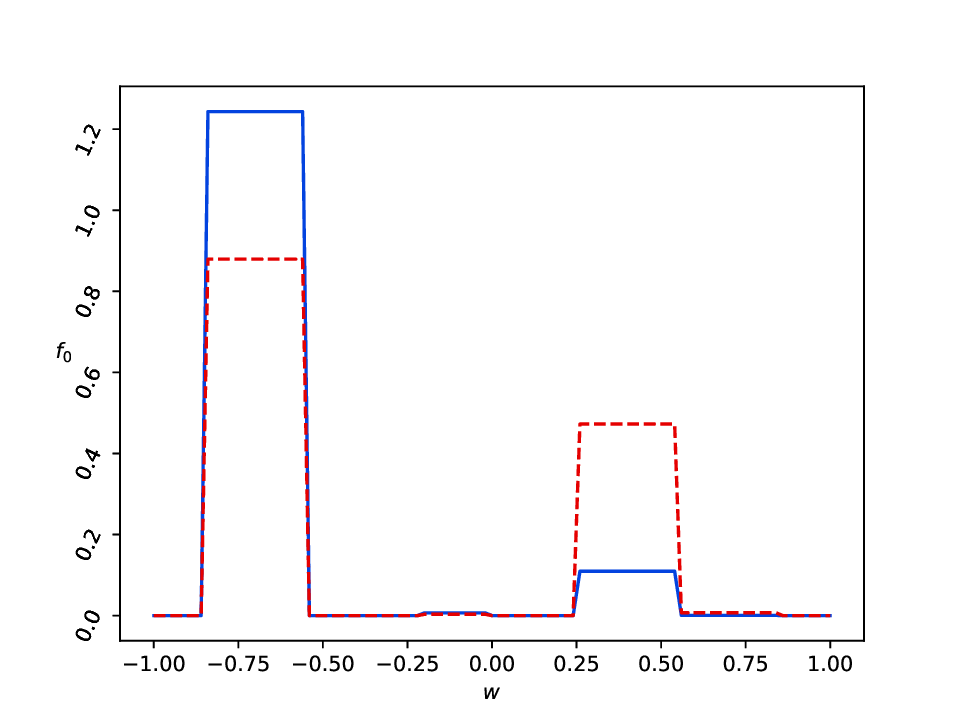}
		    \caption{18--24 year-old}
	    \end{subfigure}
     \hfill	    \begin{subfigure}[b]{0.45\linewidth}
		    \includegraphics[width=\textwidth]{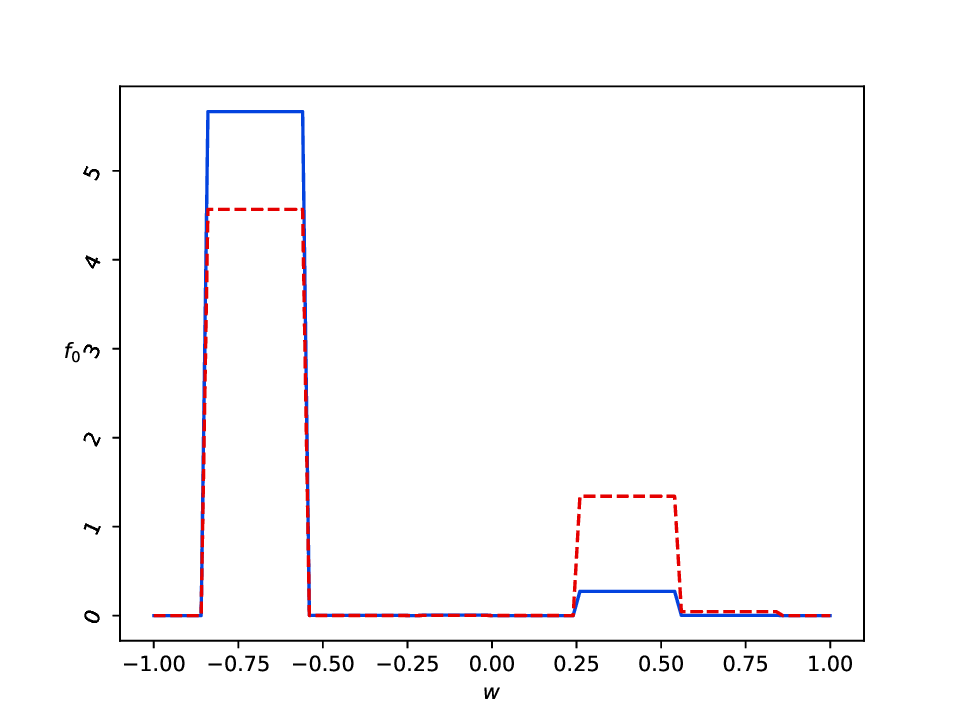}
		    \caption{25--49 year-old}
	    \end{subfigure} 
    
    	\begin{subfigure}[b]{0.45\linewidth}
	    	\includegraphics[width=\textwidth]{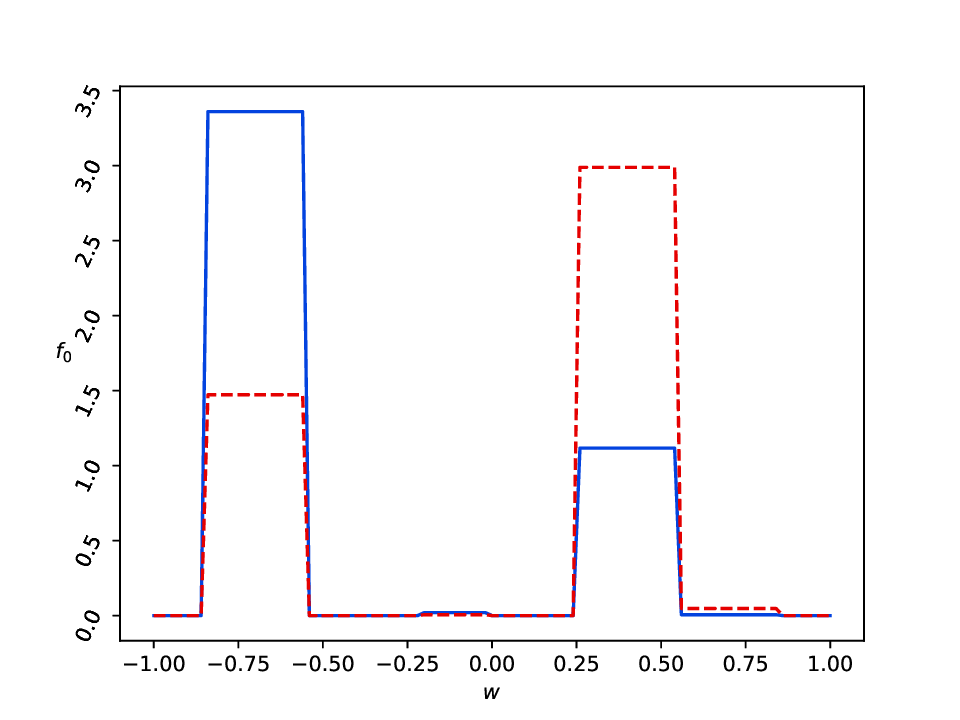}
		    \caption{50--64 year-old}
	    \end{subfigure} 
     \hfill	    \begin{subfigure}[b]{0.45\linewidth}
		    \includegraphics[width=\textwidth]{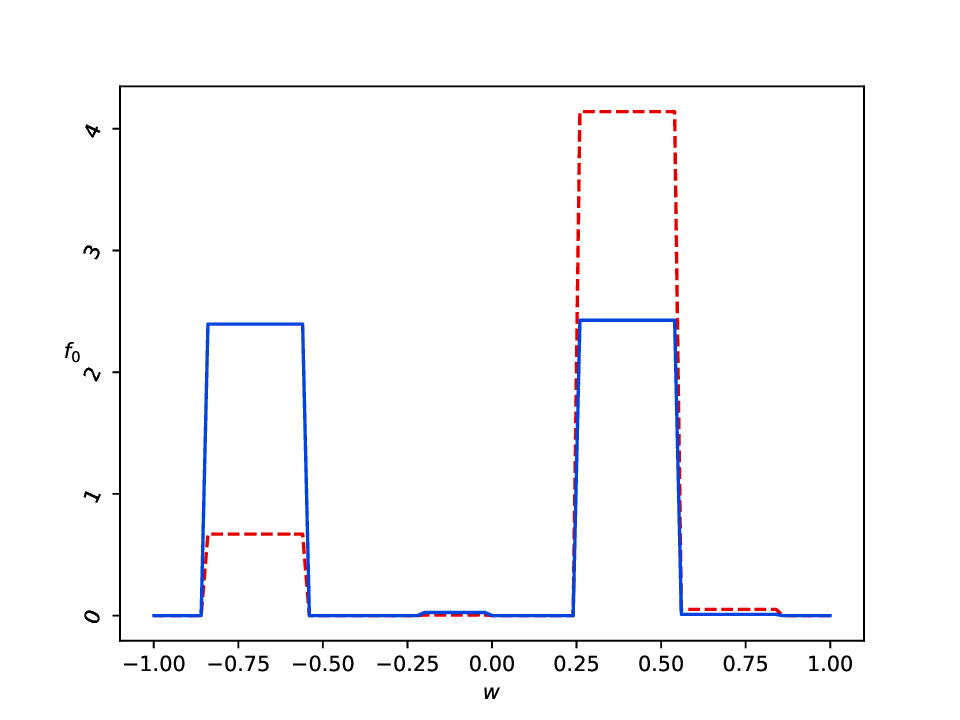}
            \caption{65+ year-old}
	    \end{subfigure}
	    \caption{Demographically motivated initial densities for the eight follower species in the constituency of Newark from the 2017 General Election Results. Each graph shows the initial density for the leave (dashed red) and the remain (solid blue) characteristic of an age group.}	
	    \label{fig:4.3analysis:newdemogICs}
    \end{figure}

    We discuss here the idea of using universal initial conditions
    against the demographic specific version used in
    \cite{bib:During:polling} (demographic initial conditions) in the
    Fokker-Planck model. Our universal initial conditions are defined
    using the raw previous election data for that constituency, that
    is when discussing $M_{\zeta, i}$ we use directly the value of
    $v_\zeta$ leaving us with the definition of $M_{\zeta,i}$
    as, $$M_{\zeta, i} = \frac{v_\zeta}{\sum_\zeta v_\zeta} M_i,$$
    with $v_\zeta$ representing the percentage of votes allocated to a
    given party, $\Pi_\zeta$, is some test election in each individual
    constituency. In this case we have chosen our test election to be
    the UK general election in 2017. This method makes the assumption
    that the vote from the 2017 general election was consistent across
    demographic. We believe this assumption is reasonable since the
    population age demographic of an individual in 2017 may be
    different from the age demographic in 2024 -- that is an
    individual who was 19 years old in 2017 will be 26 in 2024 and
    their views on political subjects may or may not have changed
    accordingly. In the previous work \cite{bib:During:polling}, we had assumed that through a reasonable usage of data, a more specific initial condition based on demographic could be synthesised. It is crucial to note at this point that in the absence of local polling data from the areas, an initial condition for this model is necessarily inaccurate. A graphical representation of the proposed universal initial conditions and the demographic initial conditions used in \cite{bib:During:polling} for the constituency of (6) Newark are presented in Figure~\ref{fig:4.3analysis:newunivICs} and Figure~\ref{fig:4.3analysis:newdemogICs}, respectively.

    \begin{figure}[h]
	    \centering
	    \begin{subfigure}[b]{0.45\linewidth} 
	        \centering
		    \includegraphics[scale=0.35]{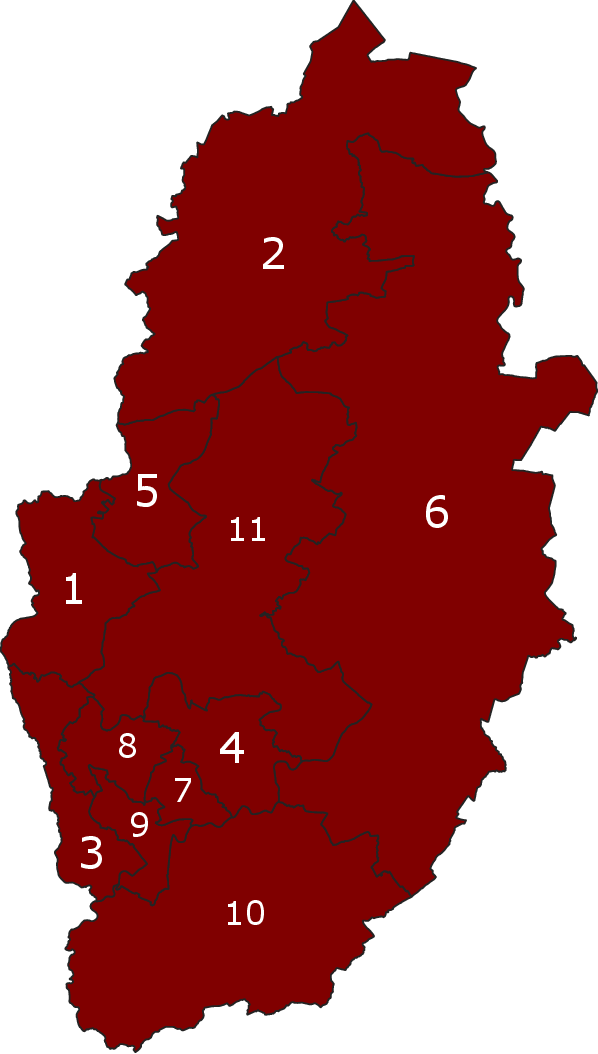}
    		\caption{Predicted Result using initial conditions as
                  in \cite{bib:During:polling}.}
    		\label{fig:4.3analysis:Not:ICBR1}
	    \end{subfigure}
     \hfill    	\begin{subfigure}[b]{0.45\linewidth} 
    	\centering
	    	\includegraphics[scale=0.35]{Not24BU1.eps}
		    \caption{Predicted Result using universal initial conditions from GE2017 results.}
		    \label{fig:4.3analysis:Not:ICBU1}
	    \end{subfigure}
    	\caption{Comparison between the different choices of initial
          conditions. Light colours (Labour in red, Conservative in blue) are where the total popular vote
          for the elected MP was below 50\% and the dark colours for greater than 50\%.}
    	\label{fig:4.3analysis:ICNot}
    \end{figure}
The results are presented in Figure~\ref{fig:4.3analysis:ICNot} with
colour-coding of each constituency, red if a Labour MP was elected,
blue if a Conservative MP was elected. The light colours are where the total popular vote for the elected MP was below 50\% and the dark colours for greater than 50\%. We can see that using the demographic initial conditions result in more extreme results -- every constituency in Nottingham is predicted a greater than 50\% vote share for Labour -- and the universal initial conditions have less extreme result -- only (7) Nottingham East, (9) South and (8) Nottingham North and Kimberley are predicted a greater than 50\% vote share for Labour. (6) Newark is of particular interest in this case because it is predicted to vote overwhelmingly Labour in Figure~\ref{fig:4.3analysis:Not:ICBR1} and slightly Conservative in Figure~\ref{fig:4.3analysis:Not:ICBU1}. The final weighted sum of vote shares from the demographics is presented in Table~\ref{table:4.3analysis:ICres}.

    \begin{table}[h] 
    \centering
    \caption{The weighted sum of the predicted vote shares from the demographics in the demographic initial conditions and universal initial conditions.}
    \label{table:4.3analysis:ICres}
    \renewcommand{\arraystretch}{1.2}
\begin{tabular}{c|cccccc} 
    Newark & Lab & Grn & LibD & Oth & Con & RefUK \\
    \hline 
    Demographic IC & 52.5\% & 7.24\% & 3.02\% & 6.00\% & 17.7\% & 13.6\% \\ 
    Universal IC & 27.7\% & 5.65\% & 4.35\% & 9.84\% & 29.8\% & 22.6\% 
\end{tabular}
    \end{table}
This effect across all constituencies is likely due to the way the demographic initial conditions incorporate the voter intention data. In order to distinguish the initial conditions in \cite{bib:During:polling}, voter intention data was multiplied together element by element and then combined with the 2015 election result. However this ``double'' usage of the voter intention data -- once in the initial conditions, once in the choice of $P$ -- makes the kinetic model susceptible to extremities in voter intention data. In the 2024 election, it has been predicted many times that Labour will win \cite{bib:YouGov:voterintention2024} and this is reflected in the voter intention data, however this likely leads to a situation with the simple demographic initial conditions, that we see above.

In subsequent experiments we will use the universal initial conditions, since this allows for a comparison of the final results without bias from the initial conditions.

\subsection{Investigating the choice of demographics}

    In previous examples we have constructed our follower species from two demographics -- namely age demographic (18-24yo, 25-49yo, 50-64yo and 65+yo) and self reported Brexit vote (leave, remain). In this experiment, we compare the contribution to predicted results from using these original follower species and eight follower species using NRS social grade (ABC1, C2DE) to replace Brexit vote. The Brexit vote had an impact on the previous election \cite{bib:BBC}, and may have an impact on the current election, however choosing to use social grade is due to its historic effect on voting in UK general elections.
        \begin{figure}[h]
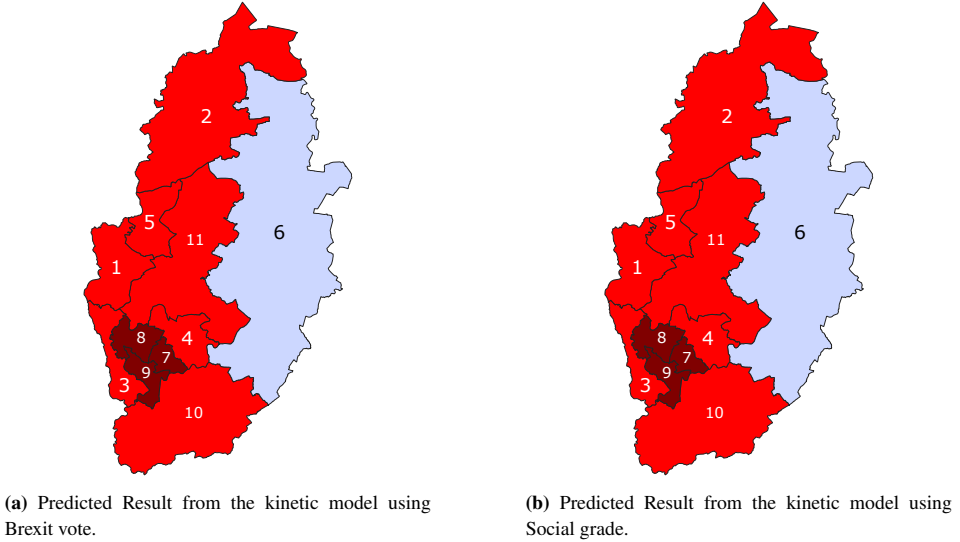

    	    \centering
    	    \begin{subfigure}[b]{0.45\linewidth} 
    	        \centering
    		    \includegraphics[scale=0.35]{Not24BU1.eps}
        		\caption{Predicted Result from the kinetic model using Brexit vote.}
        		\label{fig:4.3analysis:Not:2024brex}
    	    \end{subfigure}
     \hfill        	\begin{subfigure}[b]{0.45\linewidth} 
        	\centering
    	    	\includegraphics[scale=0.35]{Not24SU1.eps}
    		    \caption{Predicted Result from the kinetic model using Social grade.}
    		    \label{fig:4.3analysis:Not:2024social}
    	    \end{subfigure}
        	\caption{Predicted Results of the General Elections in
                  2024 in Nottinghamshire according to the model
                  presented here,\eqref{fig:4.3analysis:Not:2024brex}
                  using Brexit vote and age and,
                  \eqref{fig:4.3analysis:Not:2024social} using social
                  grade and age. The light colours  (Labour in red, Conservative in blue) are where the total popular vote for the elected MP was below 50\% and the dark colours for greater than 50\%.}
        	\label{fig:4.3analysis:demogNot}
        \end{figure}
    
    From the results presented in Figure~\ref{fig:4.3analysis:demogNot}, we can see that the party that is predicted a majority by the model does not change when changing the demographic composition of the follower species. We cannot determine whether this result will be true for all elections, since some particular choice of demographic may have some political significance -- as the Brexit vote did in \cite{bib:During:polling} -- however it appears as though these effects are less extreme in this election. Since both of these models agree in this case, we will continue to use the Brexit vote as the indicator for this election.

\subsection{Comparison to other predicted results}

Since the 2024 UK general election has been called for 4$^{th}$ July
and this work was completed before this date, we cannot compare the
model results directly to the recorded results, however we can compare
to other model predictions, such as that of YouGov
\cite{bib:YouGov:prediction2024}. Both these methods utilise the MRP
method for the prediction and use proper regionalised data, whereas we
are using national data, that has been regionalised using the method
presented in \cite{bib:During:polling}.

    \begin{figure}[h]
	    \centering
	    \begin{subfigure}[b]{0.45\linewidth} 
	        \centering
		    \includegraphics[scale=0.35]{Not24BU1.eps}
    		\caption{Predicted Result according to the kinetic model.}
    		\label{fig:4.3analysis:Not:2024us}
	    \end{subfigure}
     \hfill    	\begin{subfigure}[b]{0.45\linewidth} 
    	\centering
	    	\includegraphics[scale=0.35]{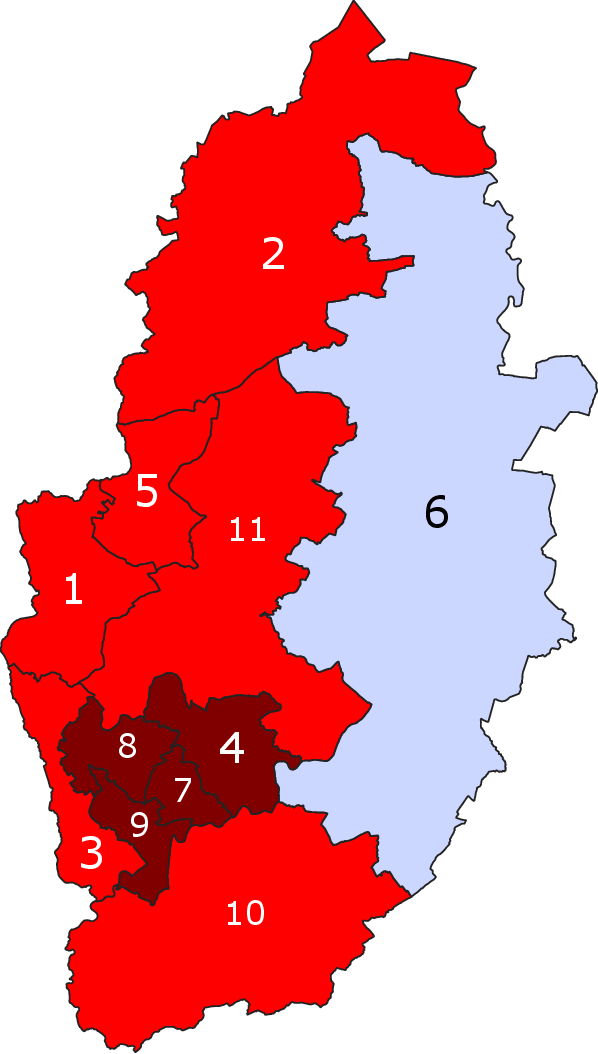}
		    \caption{Predicted Result according to the YouGov MRP model.}
		    \label{fig:4.3analysis:Not:2024YG}
	    \end{subfigure}
    	\caption{Predicted Results of the General Elections in 2024 in
          Nottinghamshire according to the model presented here, with
          $T=10$ \eqref{fig:4.3analysis:Not:2024us} and the MRP model
          presented by YouGov \eqref{fig:4.3analysis:Not:2024YG}. The light colours  (Labour in red, Conservative in blue) are where the total popular vote for the elected MP was below 50\% and the dark colours for greater than 50\%.}
    	\label{fig:4.3analysis:Not2024}
    \end{figure}

As we can see from the results in Figure~\ref{fig:4.3analysis:Not2024}, our simulation results are inline with a direct comparison with the MRP method by YouGov, with a few major differences. In terms of full results, this model predicts a less extreme win in (4) Gedling. Importantly, the secondary and tertiary results in each region, in the cases of (2) Bassetlaw, (3) Broxtowe, (4) Gedling, (5) Mansfield, (10) Rushcliffe and (11) Sherwood Forest, both the YouGov prediction and the result from the kinetic model rank at least the first three parties the same -- for all of these the ranks are 1) Labour, 2) Conservative and 3) Reform UK -- in (1) Ashfield, the Reform UK candidate, Lee Anderson, has recently defected from the Conservative party to Reform UK and is the incumbent MP in that region. We believe that for at least this case, more localised data would have been more indicative of the situation, however since we have used national data and regionalised this to a smaller area, we necessarily miss local opinions on MPs. We wish to reiterate, however, that these models agree with YouGov on which MP will have the number one ranking in each constituency, and in FPTP elections, this is the  determining statistic that informs which party will win the seat. This again demonstrates the potential use of these models for election prediction.

\subsection{A more in depth look at Bassetlaw}

We take a closer look at the constituency of Bassetlaw. We believe this constituency is of particular interest since, the seat of Bassetlaw had elected the candidate from the Labour party since 1929 with the only break being in 1931-35, where the National Labour Group -- split from the Labour party -- were incumbent. In 2019, Bassetlaw elected the candidate from the Conservative Party, Brendan Clarke-Smith, and from both the model presented here and the model presented by YouGov, the Labour Candidate is predicted to win this seat again. 

\begin{figure}
   \centering
        \includegraphics[scale=0.55]{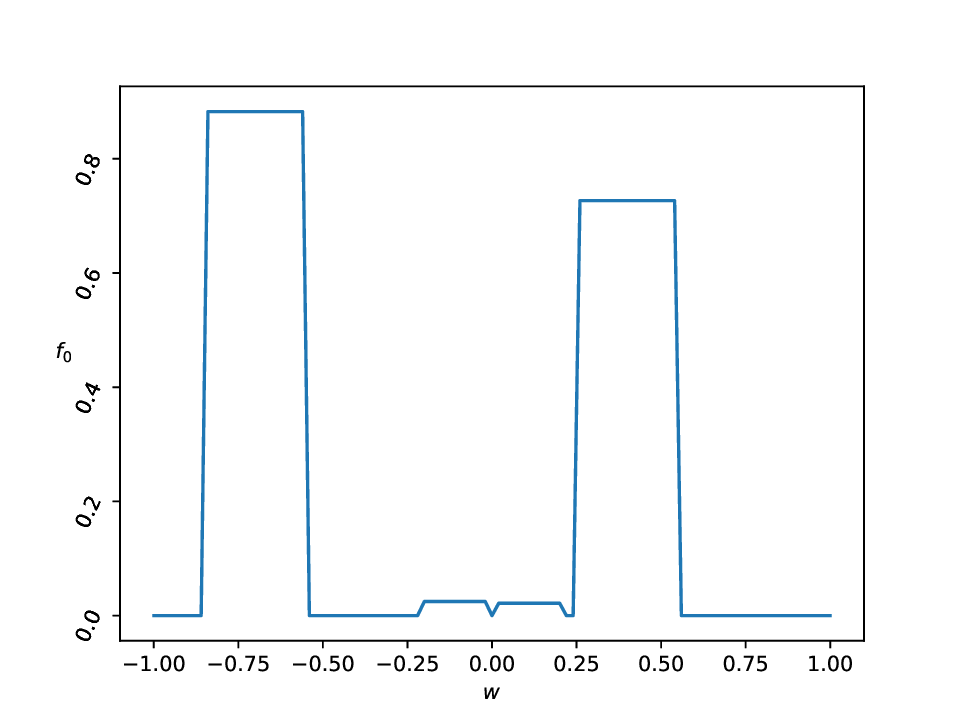}
	    \caption{The initial density for all eight follower species in Bassetlaw for the 2024 general election.}	
	    \label{chap2:sec6:BassIC}
    \end{figure}

The eight initial follower densities presented in Figure~\ref{chap2:sec6:BassIC} using universal initial conditions, from the 2017 General Election.

\begin{figure}
	    \begin{subfigure}[b]{0.45\linewidth}
		    \includegraphics[width=\textwidth]{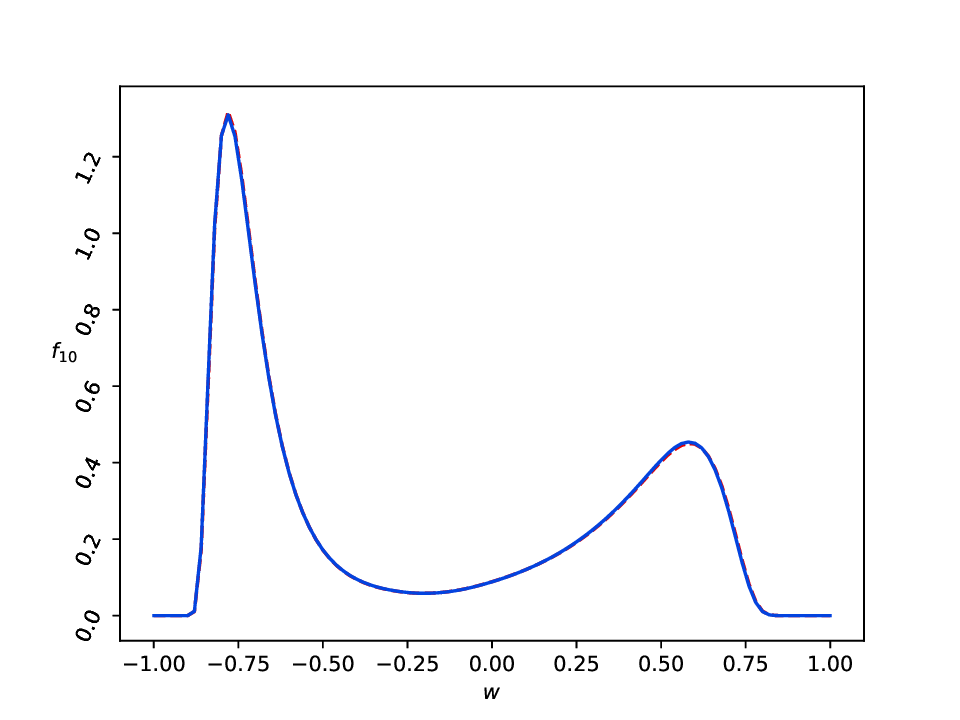}
		    \caption{18--24 year-old}
	    \end{subfigure}
		\hspace{0.04cm}
	    \begin{subfigure}[b]{0.45\linewidth}
		    \includegraphics[width=\textwidth]{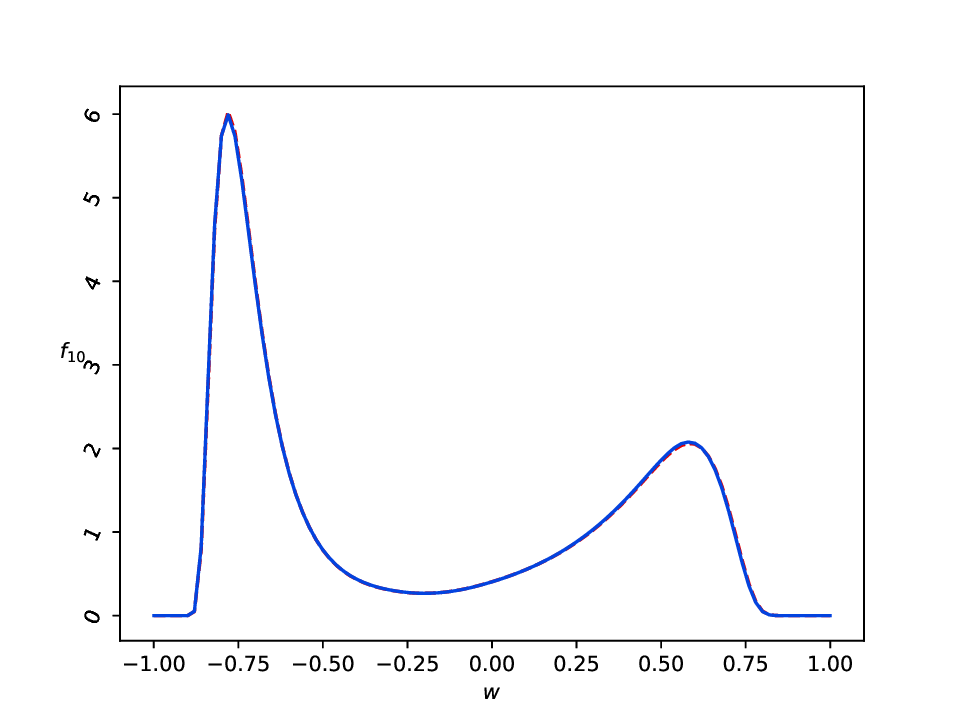}
		    \caption{25--49 year-old}
	    \end{subfigure} 
    
    	\begin{subfigure}[b]{0.45\linewidth}
	    	\includegraphics[width=\textwidth]{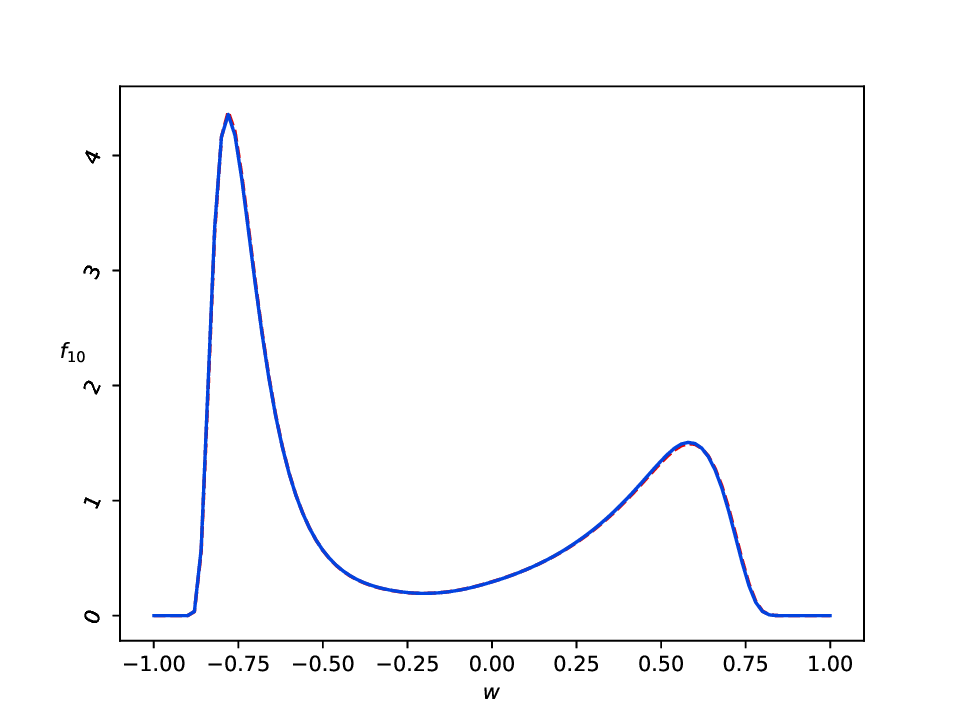}
		    \caption{50--64 year-old}
	    \end{subfigure} 
	    \hspace{0.04cm}
	    \begin{subfigure}[b]{0.45\linewidth}
		    \includegraphics[width=\textwidth]{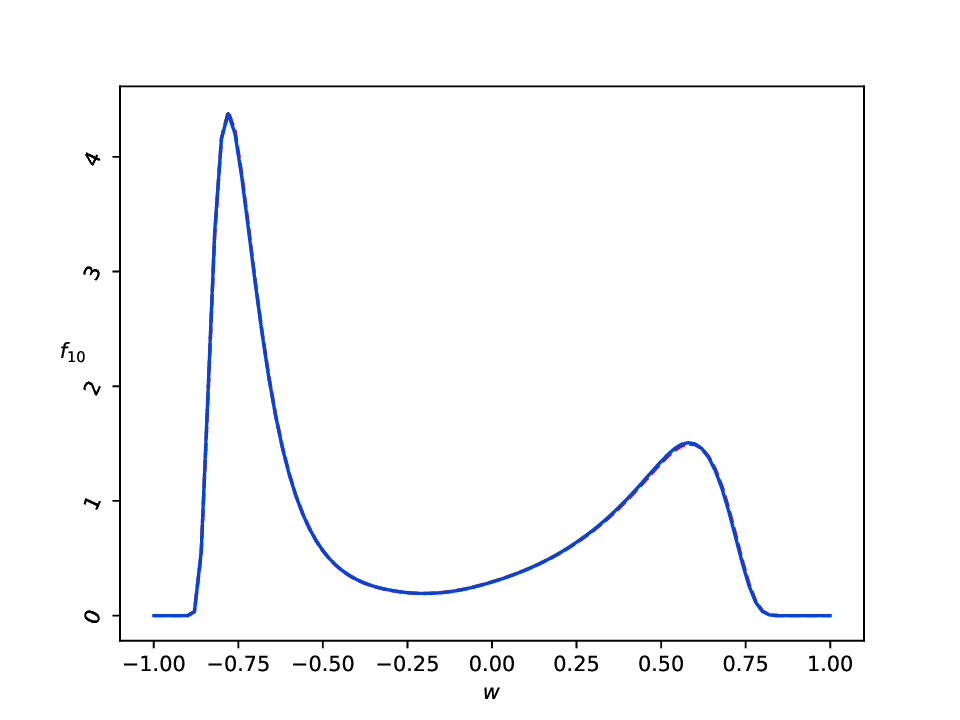 }
            \caption{65+ year-old}
	    \end{subfigure}
	    \caption{Final time  graphical representations of the predicted result from our kinetic model with $A=10$ in Bassetlaw in the 2024 General Election.}
	    \label{fig:4.3analysis:Bassres1}
    \end{figure}

As we can see from Figure~\ref{fig:4.3analysis:Bassres1}, the final time graphs are very similar for all eight follower species. The 18-24yo species have slightly more mass in the Labour influence than the 65+yo and there is a very slight separation between leave (red dotted) and remain (solid blue). In this election, YouGov and Election Calculus both predict a strong majority for Labour over the Conservatives, and this is especially strong in the Red Wall. These graphs will be different if we were to use true local polling data, however the collated result in Bassetlaw is very similar to that of YouGov and Electoral Calculus.

\begin{figure}
	    \begin{subfigure}[b]{0.45\linewidth}
		    \includegraphics[width=\textwidth]{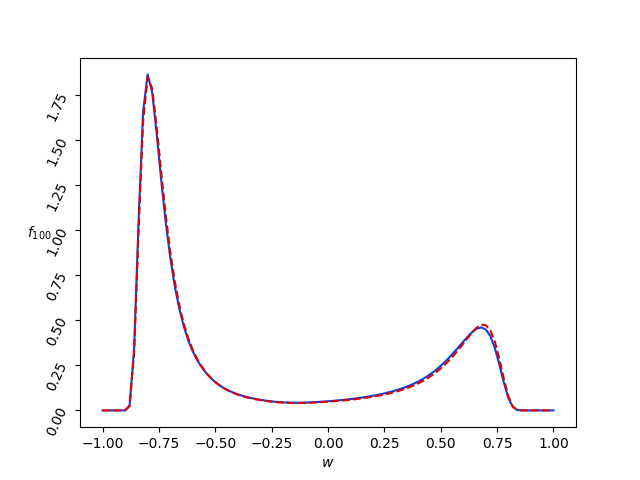}
		    \caption{18--24 year-old}
	    \end{subfigure}
		\hspace{0.04cm}
	    \begin{subfigure}[b]{0.45\linewidth}
		    \includegraphics[width=\textwidth]{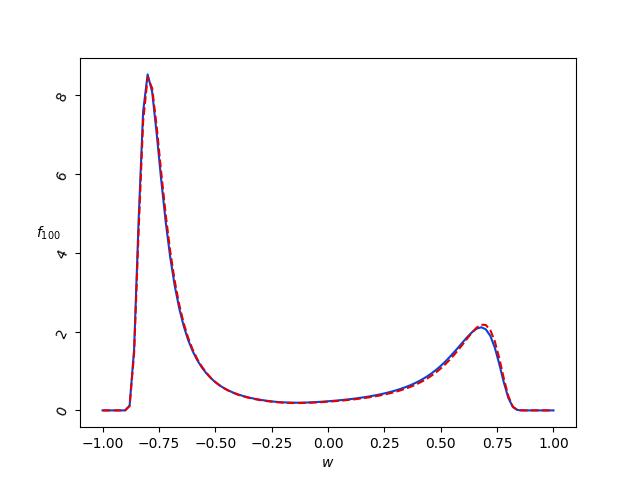}
		    \caption{25--49 year-old}
	    \end{subfigure} 
    
    	\begin{subfigure}[b]{0.45\linewidth}
	    	\includegraphics[width=\textwidth]{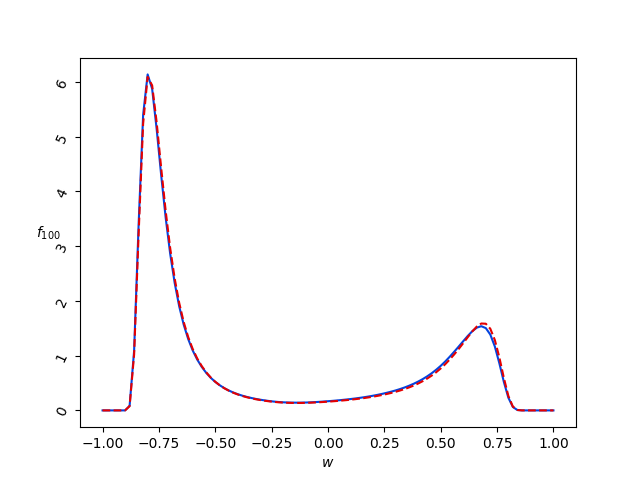}
		    \caption{50--64 year-old}
	    \end{subfigure} 
	    \hspace{0.04cm}
	    \begin{subfigure}[b]{0.45\linewidth}
		    \includegraphics[width=\textwidth]{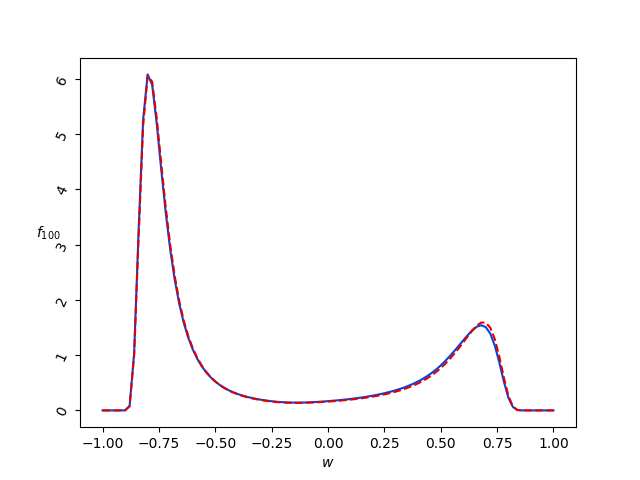}
            \caption{65+ year-old}
	    \end{subfigure}
	    \caption{Final time  graphical representations of the predicted result from our kinetic model with $A=100$ in Bassetlaw in the 2024 General Election.}
	    \label{fig:4.3analysis:Bassres10}
    \end{figure}

We also wish to compare the results in Figure~\ref{fig:4.3analysis:Bassres1}, to the results obtained using $A=100$, Figure~\ref{fig:4.3analysis:Bassres10}, since this clearly shows a lack of differentiation between the follower species in Bassetlaw, even when looking at substantially longer time steps. This confirms that our prediction is we should see a strong Labour win in Bassetlaw.

\begin{funding}
This work was partially supported by a Royal Society International
Ex\-chan\-ges grant (Ref. IES/R3/213113). O.W.\ is supported by the Warwick
Mathematics Institute Centre for Doctoral Training in Mathematics, and gratefully acknowledges funding from the University of Warwick.
\end{funding}


\bibliographystyle{ems}
\bibliography{bdbib}

\begin{thebibliography}{10}
\providecommand{\url}[1]{\texttt{#1}}
\providecommand{\urlprefix}{URL }
\providecommand{\eprint}[2][]{\url{#2}}

\bibitem{albi2024data}
G.~Albi, E.~Calzola, and G.~Dimarco,
  \href{https://doi.org/10.1017/S0956792524000068}{A data-driven kinetic model
  for opinion dynamics with social network contacts}. \emph{European Journal of
  Applied Mathematics}  (2024), 1--27

\bibitem{albi2016recent}
G.~Albi, L.~Pareschi, G.~Toscani, and M.~Zanella,
  \href{https://doi.org/10.1007/978-3-319-49996-3_2}{Recent advances in opinion
  modeling: control and social influence}. In \emph{{Active Particles, Volume
  1. Modeling and Simulation in Science, Engineering and Technology}}, edited
  by N.~Bellomo, P.~Degond, and E.~Tadmor, pp. 49--98, Birkhäuser, Cham, 2017

\bibitem{albi2015optimal}
G.~Albi, L.~Pareschi, and M.~Zanella,
  \href{https://doi.org/10.1007/978-3-319-55795-3_4}{On the optimal control of
  opinion dynamics on evolving networks}. In \emph{System modeling and
  optimization}, edited by L.~Bociu, J.-A. D{\'e}sid{\'e}ri, and A.~Habbal, pp.
  58--67, Springer, Cham, 2016

\bibitem{albi2016opinion}
G.~Albi, L.~Pareschi, and M.~Zanella,
  \href{https://doi.org/10.3934/krm.2017001}{Opinion dynamics over complex
  networks: kinetic modeling and numerical methods}. \emph{Kinet. Relat.
  Models} \textbf{10} (2017), 1--32

\bibitem{albi2019boltzmann}
G.~Albi, L.~Pareschi, and M.~Zanella,
  \href{https://doi.org/10.1007/s10955-019-02246-y}{Boltzmann games in
  heterogeneous consensus dynamics}. \emph{J. Stat. Phys.} \textbf{175} (2019),
  no.~1, 97--125

\bibitem{bib:BBC}
{BBC News}, General election 2019: {C}onservatives take {W}akefield from
  {L}abour. 2019,
  \urlprefix\url{https://www.bbc.co.uk/news/election-2019-50772224}, last
  accessed 1 July 2021

\bibitem{zbMATH06553232}
N.~{Bellomo}, F.~{Colasuonno}, D.~{Knopoff}, and J.~{Soler},
  \href{https://doi.org/10.3934/nhm.2015.10.421}{{From a systems theory of
  sociology to modeling the onset and evolution of criminality}}. \emph{{Netw.
  Heterog. Media}} \textbf{10} (2015), no.~3, 421--441

\bibitem{bertotti2008discrete}
M.~L. Bertotti and M.~Delitala,
  \href{https://doi.org/10.1016/j.mcm.2007.12.021}{On a discrete generalized
  kinetic approach for modelling persuader’s influence in opinion formation
  processes}. \emph{Math. comput. model.} \textbf{48} (2008), no. 7-8,
  1107--1121

\bibitem{zbMATH06977672}
L.~{Boudin} and F.~{Salvarani},
  \href{https://doi.org/10.1016/j.physa.2015.10.014}{{Opinion dynamics: kinetic
  modelling with mass media, application to the Scottish independence
  referendum}}. \emph{{Physica A}} \textbf{444} (2016), 448--457

\bibitem{burger2021network}
M.~Burger, \href{https://doi.org/10.1007/s10013-021-00505-8}{Network structured
  kinetic models of social interactions}. \emph{Vietnam J. Math.}  (2021)

\bibitem{zbMATH06892489}
M.~{Burger}, M.~{Di Francesco}, S.~{Fagioli}, and A.~{Stevens},
  \href{https://doi.org/10.1137/17M1125716}{{Sorting phenomena in a
  mathematical model for two mutually attracting/repelling species}}.
  \emph{{SIAM J. Math. Anal.}} \textbf{50} (2018), no.~3, 3210--3250

\bibitem{zbMATH06712347}
B.~{Chazelle}, Q.~{Jiu}, Q.~{Li}, and C.~{Wang},
  \href{https://doi.org/10.1016/j.jde.2017.02.036}{{Well-posedness of the
  limiting equation of a noisy consensus model in opinion dynamics}}. \emph{{J.
  Differ. Equations}} \textbf{263} (2017), no.~1, 365--397

\bibitem{zbMATH06875734}
E.~{Cristiani} and A.~{Tosin},
  \href{https://doi.org/10.1137/17M113397X}{{Reducing complexity of multiagent
  systems with symmetry breaking: an application to opinion dynamics with
  polls}}. \emph{{Multiscale Model. Simul.}} \textbf{16} (2018), no.~1,
  528--549

\bibitem{zbMATH06684856}
P.~{Degond}, J.-G. {Liu}, S.~{Merino-Aceituno}, and T.~{Tardiveau},
  \href{https://doi.org/10.1142/S021820251740005X}{{Continuum dynamics of the
  intention field under weakly cohesive social interaction}}. \emph{{Math.
  Models Methods Appl. Sci.}} \textbf{27} (2017), no.~1, 159--182

\bibitem{during2024breaking}
B.~D{\"u}ring, J.~Franceschi, M.-T. Wolfram, and M.~Zanella,
  \href{https://doi.org/10.1007/s00332-024-10060-4}{Breaking consensus in
  kinetic opinion formation models on graphons}. \emph{Journal of Nonlinear
  Science} \textbf{34} (2024), no.~4, 79

\bibitem{zbMATH06665924}
B.~{D\"uring}, A.~{J\"ungel}, and L.~{Trussardi},
  \href{https://doi.org/10.3934/krm.2017010}{{A kinetic equation for economic
  value estimation with irrationality and herding}}. \emph{{Kinet. Relat.
  Models}} \textbf{10} (2017), no.~1, 239--261

\bibitem{bib:During:economykineticmodel}
B.~D\"uring, D.~Matthes, and G.~Toscani,
  \href{https://doi.org/10.1103/PhysRevE.78.056103}{Kinetic equations modelling
  wealth redistribution: A comparison of approaches}. \emph{Phys. Rev. E}
  \textbf{78} (2008), 056103

\bibitem{bib:During:strongleaders}
B.~Düring, P.~Markowich, J.~Pietschmann, and M.~Wolfram,
  \href{https://doi.org/10.1098/rspa.2009.0239}{{B}oltzmann and
  {F}okker-{P}lanck equations modelling opinion formation in the presence of
  strong leaders}. \emph{Proc. R. Soc. A.} \textbf{465} (2009), 3687--3708

\bibitem{bib:During:inhomogeneous}
B.~Düring and M.~Wolfram,
  \href{https://doi.org/10.1098/rspa.2015.0345}{Opinion dynamics: inhomogeneous
  {B}oltzmann-type equations modelling opinion leadership and political
  segregation}. \emph{Proc. R. Soc. A.} \textbf{471} (2015), 20150345

\bibitem{bib:During:polling}
B.~Düring and O.~Wright, \href{https://doi.org/10.1098/rsta.2021.0154}{On a
  kinetic opinion formation model for pre-election polling}. \emph{Phil. Trans.
  R. Soc. A.} \textbf{380} (2022), no. 2224, 20210154

\bibitem{zbMATH06684855}
G.~{Furioli}, A.~{Pulvirenti}, E.~{Terraneo}, and G.~{Toscani},
  \href{https://doi.org/10.1142/S0218202517400048}{{Fokker-Planck equations in
  the modeling of socio-economic phenomena}}. \emph{{Math. Models Methods Appl.
  Sci.}} \textbf{27} (2017), no.~1, 115--158

\bibitem{zbMATH06735322}
J.~{Garnier}, G.~{Papanicolaou}, and T.-W. {Yang},
  \href{https://doi.org/10.1007/s10013-016-0190-2}{{Consensus convergence with
  stochastic effects}}. \emph{{Vietnam J. Math.}} \textbf{45} (2017), no. 1-2,
  51--75

\bibitem{bib:HOC:2017GE}
{House of Commons Library}, Data file: detailed results by constituency
  [microsoft excel spreadsheet]. 2019,
  \urlprefix\url{https://commonslibrary.parliament.uk/research-briefings/cbp-7979},
  last accessed 5 March 2021

\bibitem{bib:HOC:2019GE}
{House of Commons Library}, Data file: detailed results by constituency
  [microsoft excel spreadsheet]. 2021,
  \urlprefix\url{https://commonslibrary.parliament.uk/research-briefings/cbp-8749},
  last accessed 9 June 2021

\bibitem{zbMATH06350891}
P.-E. {Jabin} and S.~{Motsch},
  \href{https://doi.org/10.1016/j.jde.2014.08.005}{{Clustering and asymptotic
  behavior in opinion formation}}. \emph{{J. Differ. Equations}} \textbf{257}
  (2014), no.~11, 4165--4187

\bibitem{motsch2014heterophilious}
S.~Motsch and E.~Tadmor,
  \href{https://doi.org/10.1137/120901866}{Heterophilious dynamics enhances
  consensus}. \emph{SIAM Rev.} \textbf{56} (2014), no.~4, 577--621

\bibitem{bib:ONS:2011census}
{Office of National Statistics}, 2011 census, {KS102EW} age structure, local
  authorities in {E}ngland and {W}ales. 2012,
  \urlprefix\url{http://www.nomisweb.co.uk/census/2011/ks102ew}, last accessed
  1 March 2021.

\bibitem{zbMATH07146312}
L.~{Pareschi}, G.~{Toscani}, A.~{Tosin}, and M.~{Zanella},
  \href{https://doi.org/10.1007/s00332-019-09558-z}{{Hydrodynamic models of
  preference formation in multi-agent societies}}. \emph{{J. Nonlinear Sci.}}
  \textbf{29} (2019), no.~6, 2761--2796

\bibitem{park2004bayesian}
D.~K. Park, A.~Gelman, and J.~Bafumi,
  \href{https://doi.org/10.1093/pan/mph024}{Bayesian multilevel estimation with
  poststratification: State-level estimates from national polls}.
  \emph{Political Anal.} \textbf{12} (2004), 375--385

\bibitem{zbMATH07205672}
L.~{Pedraza}, J.~P. {Pinasco}, and N.~{Saintier},
  \href{https://doi.org/10.1142/S0218202520500062}{{Measure-valued opinion
  dynamics}}. \emph{{Math. Models Methods Appl. Sci.}} \textbf{30} (2020),
  no.~2, 225--260

\bibitem{zbMATH06944946}
M.~{P\'erez-Llanos}, J.~P. {Pinasco}, N.~{Saintier}, and A.~{Silva},
  \href{https://doi.org/10.1137/17M1152784}{{Opinion formation models with
  heterogeneous persuasion and zealotry}}. \emph{{SIAM J. Math. Anal.}}
  \textbf{50} (2018), no.~5, 4812--4837

\bibitem{bib:ElectoralComission:2016Brexit}
{The Electoral Commission}, Results and turnout at the {EU} referendum. 2019,
  \urlprefix\url{https://www.electoralcommission.org.uk/who-we-are-and-what-we-do/elections-and-referendums/past-elections-and-referendums/eu-referendum/results-and-turnout-eu-referendum},
  last accessed 16 April 2021

\bibitem{bib:Toscani:kineticmodel}
G.~Toscani, \href{https://doi.org/10.4310/CMS.2006.v4.n3.a1}{Kinetic models of
  opinion formation}. \emph{Comm. Math. Sci.} \textbf{4} (2006), no.~3,
  481--496

\bibitem{bib:ToscaniPareschi:ims}
G.~Toscani and L.~Pareschi, \emph{{Interacting Multiagent Systems}}. Oxford
  University Press, 2013

\bibitem{toscani2018opinion}
G.~Toscani, A.~Tosin, and M.~Zanella,
  \href{https://doi.org/10.1103/PhysRevE.98.022315}{Opinion modeling on social
  media and marketing aspects}. \emph{Phys. Rev. E} \textbf{98} (2018), no.~2,
  022315

\bibitem{bib:YouGov:prediction2024}
YouGov, {UK} general election 2024. 2024,
  \urlprefix\url{https://yougov.co.uk/elections/uk/2024}, last accessed 29 June
  2024

\bibitem{bib:YouGov:voterintention2024}
YouGov, Voting intention. 2024,
  \urlprefix\url{https://yougov.co.uk/topics/politics/trackers/voting-intention},
  last accessed 29 June 2024

\bibitem{zanella2023kinetic}
M.~Zanella, \href{https://doi.org/10.1007/s11538-023-01147-2}{Kinetic models
  for epidemic dynamics in the presence of opinion polarization}.
  \emph{Bulletin of Mathematical Biology} \textbf{85} (2023), no.~5, 36

\end{thebibliography}

\end{document}